\documentclass[twocolumn,prl,superscriptaddress]{revtex4}[10pt]
\usepackage{graphicx}

\begin{document}

\def\simgt{\mathrel{\lower2.5pt\vbox{\lineskip=0pt\baselineskip=0pt
           \hbox{$>$}\hbox{$\sim$}}}}
\def\simlt{\mathrel{\lower2.5pt\vbox{\lineskip=0pt\baselineskip=0pt
           \hbox{$<$}\hbox{$\sim$}}}}
\newcommand{\met}{\mbox{${\rm \not\! E}_{\rm T}$}}

\title{A New Approach to {\boldmath $\mu\mbox{-}B_\mu$}}

\author{Csaba Cs\'aki}
\affiliation{Institute for High Energy Phenomenology,
 Newman Laboratory of Elementary Particle Physics,
 Cornell University, Ithaca, NY 14853, USA}

\author{Adam Falkowski}
\affiliation{CERN, Theory Division, CH-1211 Geneva 23, Switzerland}

\author{Yasunori Nomura}
\affiliation{Department of Physics, University of California,
 Berkeley, CA 94720, USA}
\affiliation{Theoretical Physics Group, Lawrence Berkeley National Laboratory,
 Berkeley, CA 94720, USA}

\author{Tomer Volansky}
\affiliation{ School of Natural Sciences, Institute for Advanced Study,
 Princeton, NJ 08540}

\begin{abstract}
We present a new approach to the $\mu\mbox{-}B_\mu$ problem of gauge 
mediated supersymmetry breaking.  Rather than reducing the generically 
large contribution to $B_\mu$ we point out that acceptable electroweak 
symmetry breaking can be achieved with $\mu^2 \ll B_\mu$ if at the same 
time $B_\mu \ll m_{H_d}^2$.  This hierarchy can easily appear in models 
where the Higgs fields are directly coupled to the supersymmetry breaking 
sector.  Such models can yield novel electroweak symmetry breaking vacua, 
can deal with the supersymmetric flavor and $CP$ problems, allow for 
gauge coupling unification, and result in distinct phenomenological 
predictions for the spectrum of superparticles.
\end{abstract}

\maketitle

{\bf Introduction.}
Supersymmetry (SUSY) is a very attractive candidate for explaining the 
stability of the weak scale against radiative corrections.  However, 
it suffers from nagging problems such as the SUSY flavor problem, the 
SUSY $CP$ problem, and the $\mu$ problem.  The SUSY flavor problem 
points toward gauge mediated SUSY breaking (GMSB)~\cite{Dine:1981gu}. 
However, GMSB itself suffers from the variant of the $\mu$ problem called 
the $\mu\mbox{-}B_\mu$ problem~\cite{Dvali:1996cu}.  The problem lies 
in the fact that generic GMSB models predict the relation $B_\mu \approx 
16\pi^2 \mu^2 \gg \mu^2$ which prevents electroweak symmetry breaking 
(EWSB) if the soft masses in the Higgs sector are of the same order 
as $\mu$.  Typically, solving the $\mu\mbox{-}B_\mu$ problem is achieved 
by introducing some additional dynamics that ensures $B_\mu \simlt 
\mu^2$ (see e.g.~\cite{Dvali:1996cu,Langacker:1999hs,Dine:1997qj}).

In this letter we propose a new approach to the $\mu\mbox{-}B_\mu$ problem. 
We point out that the GMSB relation $B_\mu \gg \mu^2$ does not pose 
any problem if we allow the other mass parameters in the Higgs sector 
to also display a hierarchy.  In particular, we argue that the pattern
\begin{equation}
  \mu^2 \sim m_{H_u}^2 \ll B_\mu \ll m_{H_d}^2,
\label{eq:pattern}
\end{equation}
leads to viable EWSB.  We show that this hierarchical pattern with 
the down-type Higgs soft mass dominating over $B_\mu$ can be naturally 
obtained in GMSB models where the Higgs multiplets are directly coupled 
to the SUSY breaking sector.  Finally, we argue that the pattern in 
Eq.~(\ref{eq:pattern}) leads to interesting novel phenomenology. 
Many more details of this scenario will be discussed in~\cite{CFNV}.

{\bf The {\boldmath $\mu\mbox{-}B_\mu$} problem.}
Let us begin by reviewing the $\mu\mbox{-}B_\mu$ problem of GMSB.  The 
tree level equations for the Higgs vacuum expectation values are given by
\begin{eqnarray}
  \frac{m_Z^2}{2} &=&
    - |\mu|^2 - \frac{m_{H_u}^2 \tan^2\!\beta - m_{H_d}^2}{\tan^2\!\beta-1},
\label{eq:minim-1}\\
  \sin 2\beta &=& \frac{2 B_\mu}{2|\mu|^2 + m_{H_u}^2 + m_{H_d}^2}.
\label{eq:minim-2}
\end{eqnarray}
Here we adopt the convention $B_\mu > 0$.  The first of these equations 
represents a problematic aspect of the minimal supersymmetric standard 
model (MSSM) known as the $\mu$ problem: the SUSY preserving parameter 
$\mu$ is required to be related to the SUSY breaking masses and, in the 
absence of fine-tuning, both should be of the order of the weak scale. 
A solution to the problem can arise if $\mu$ is generated in conjunction 
with SUSY breaking.  In the limit of $\mu = 0$, the MSSM has an 
enhanced Peccei-Quinn (PQ) symmetry.  If SUSY breaking leads to the 
breaking of this (accidental) symmetry, a $\mu$ parameter of the 
correct magnitude can be generated.  While this idea can be elegantly 
realized in the context of gravity mediation~\cite{Giudice:1988yz}, 
it encounters a problem in the framework of GMSB~\cite{Dvali:1996cu}. 
In order to dynamically generate $\mu$ of order the SUSY breaking masses 
in GMSB, the PQ symmetry must be broken by coupling the Higgs fields 
directly to the SUSY breaking sector.  Such couplings, however, typically 
generate both $\mu$ and $B_\mu$ at one loop, leading to the relation 
$B_\mu \approx 16\pi^2 \mu^2 \gg \mu^2$.  This is said to be problematic 
for the following reason.  If $m_{H_u}^2 \sim m_{H_d}^2 \sim \mu^2$, as 
suggested naively by Eq.~(\ref{eq:minim-1}), then Eq.~(\ref{eq:minim-2}) 
cannot be satisfied with $B_\mu \gg \mu^2$.  If, on the other hand, 
$m_{H_u}^2 \sim m_{H_d}^2 \sim B_\mu$, significant fine-tuning is needed 
to satisfy Eq.~(\ref{eq:minim-1}), since the experimental constraint of 
$\mu \simgt m_Z$ then requires $m_{H_u}^2, m_{H_d}^2, B_\mu$ to be much 
larger than $m_Z^2$.  This is the notorious $\mu\mbox{-}B_\mu$ problem, 
which is considered to be one of the most serious problems of GMSB.

{\bf Basic proposal.}
The common lore is that a solution to the $\mu\mbox{-}B_\mu$ problem 
must reduce the hierarchy between $\mu$ and $B_\mu$.  A close inspection 
of Eqs.~(\ref{eq:minim-1},~\ref{eq:minim-2}) reveals, however, that this 
is not necessary.  To have a solution to Eq.~(\ref{eq:minim-2}), it 
is sufficient that $2 B_\mu < m_{H_u}^2 + m_{H_d}^2 + 2|\mu|^2$.  This 
suggests a completely different approach to the $\mu\mbox{-}B_\mu$ 
problem: we can keep the generic GMSB relation $B_\mu \gg \mu^2$ and, 
at the same time, generate the hierarchy $m_{H_d}^2 \gg B_\mu$ together 
with $m_{H_u}^2 \sim \mu^2$.

To be more specific, let us consider the following simple pattern of 
the mass parameters in the Higgs sector:
\begin{equation}
  \mu \approx \epsilon \Lambda_H,
\
  B_\mu \approx \epsilon \Lambda_H^2,
\
  m_{H_u}^2 \approx \epsilon^2 \Lambda_H^2,
\
  m_{H_d}^2 \approx \Lambda_H^2,
\label{eq:muBmu-naive}
\end{equation}
where $\epsilon \ll 1$ and $\Lambda_H$ is the effective SUSY breaking 
scale in the Higgs sector.  It is easy to see that the pattern of 
Eq.~(\ref{eq:muBmu-naive}) leads to a correct EWSB vacuum.  The EWSB 
stability condition $2|\mu|^2 + m_{H_u}^2 + m_{H_d}^2 > 2 B_\mu$ is 
satisfied for $m_{H_d}^2 > 0$, since $m_{H_d}^2$ is parametrically 
larger than $B_\mu$.  Consequently, Eq.~(\ref{eq:minim-2}) can be 
solved with
\begin{equation}
  \tan\beta \approx \frac{m_{H_d}^2}{B_\mu} \approx \frac{1}{\epsilon}.
\label{eq:tanb_naive}
\end{equation}
The condition for EWSB is $B_\mu^2 > (|\mu|^2 + m_{H_u}^2)(|\mu|^2 
+ m_{H_d}^2)$.  In a typical SUSY breaking scenario, this condition 
is fulfilled as a result of renormalization group evolution making 
$m_{H_u}^2$ negative.  While this could also happen in our scenario, 
it is not necessary.  Since $B_\mu^2 \approx \mu^2 m_{H_d}^2 
\approx m_{H_u}^2 m_{H_d}^2$, both sides of the above inequality 
are parametrically of the same order of magnitude, and therefore the 
condition can be satisfied with positive $m_{H_u}^2$.  The negative 
radiative corrections from top-stop loops are of course still present, 
but they do not have to be dominant.  EWSB is then driven by the dynamics 
that generate the boundary conditions, Eq.~(\ref{eq:muBmu-naive}), 
rather than by the renormalization group (RG) evolution of the MSSM 
(see also~\cite{Nomura:2005rk} for a related idea).

Let us now look at the vacuum equation~(\ref{eq:minim-1}).  The 
three contributions to the right-hand-side (RHS) are all of the 
same order $\mu^2 \approx m_{H_u}^2 \approx m_{H_d}^2/\tan^2\!\beta 
\approx \epsilon^2 \Lambda_H^2$.  This suggests that the pattern of 
Eq.~(\ref{eq:muBmu-naive}) could potentially lead to a ``fully natural'' 
EWSB: if $\epsilon^2 \Lambda_H^2 \approx m_Z^2$ the electroweak symmetry 
can be broken without any fine-tuning.  Unfortunately, a careful analysis 
shows that the situation is not that simple, and we still likely need 
some amount of cancellations in the RHS of Eq.~(\ref{eq:minim-1}) in 
realistic parameter regions.  The reason is that satisfying the LEP~II 
bound on the Higgs boson mass requires a rather heavy stop, $m_{\tilde{t}} 
\simgt 1~{\rm TeV}$, which then feeds in as a large one-loop contribution 
of order $(500~{\rm GeV})^2$ to $m_{H_u}^2$.  One should be aware of 
the fact that if this ``irreducible'' fine-tuning is not eliminated 
there is no compelling motivation to consider the ``fully natural'' 
pattern of Eq.~(\ref{eq:muBmu-naive}).  Instead, one can consider 
more general scenarios for the Higgs mass parameters which still lead 
to realistic EWSB without improving but not worsening the fine-tuning. 
In fact, theories discussed below can also lead to such a generalization 
of Eq.~(\ref{eq:muBmu-naive}).

The ``irreducible'' fine-tuning described above may be ameliorated 
if we go beyond the MSSM with simple GMSB soft terms.  Improving the 
situation requires extra contributions to the Higgs quartic coupling 
and/or to the scalar trilinear couplings.  Such contributions may 
appear due to direct couplings of the Higgs fields to the SUSY breaking 
sector, or with the aid of extra singlet fields.  The possibility of 
raising $m_{H_d}^2$ with $B_\mu \approx \mu^2$ was considered in the 
context of NMSSM-type models in~\cite{Dine:1997qj}.

{\bf Realization.}
Our approach to the $\mu\mbox{-}B_\mu$ problem requires some dynamics 
that naturally generates the pattern of Eq.~(\ref{eq:muBmu-naive}) 
or its variants.  Moreover, a natural theory should relate the scale 
$\Lambda_H$ to that for the gaugino, squark and slepton masses.  In 
the following we explain how to achieve these by coupling the MSSM 
Higgs fields directly to the SUSY breaking sector.

First, recall general features for the soft gaugino masses $M_a$ and 
scalar masses $m_I^2$ in the GMSB framework.  Below we define our 
parameters at the scale $M$ where SUSY breaking effects are mediated 
to the MSSM sector.  In perturbative gauge mediation, $M$ corresponds 
to the mass scale for the messenger fields.  Unless there is a special 
structure in the SUSY breaking sector (e.g.~an approximate $R$ symmetry), 
the soft masses take the form:
\begin{equation}
  M_a \approx g_a^2 \frac{N}{16\pi^2} \Lambda,
\quad
  m_I^2 \approx
    \sum_{a=1,2,3} \frac{g_a^4 C_I^a}{8\pi^2}
    \frac{N}{16\pi^2} \Lambda^2.
\label{eq:GMSB}
\end{equation}
Here, $g_a$ are the MSSM gauge couplings evaluated at $M$, and $C_I^a$ 
are the quadratic Casimir coefficients.  The quantity $N$ measures 
the number of SUSY breaking sector fields charged under the MSSM 
gauge group, and $\Lambda$ is the effective SUSY breaking scale. 
In perturbative gauge mediation, $\Lambda$ is the SUSY breaking 
mass squared splitting divided by the SUSY mass for the messenger 
fields, $F/M$.  Note that, in general, there can be ${\cal O}(1)$ 
coefficients in the RHSs of Eq.~(\ref{eq:GMSB}).  For general 
expressions for $M_a$ and $m_I^2$ in GMSB, see~\cite{Meade:2008wd}.

In order to generate $\mu$ and $B_\mu$, we consider the superpotential 
couplings of the Higgs fields $H_{u,d}$ to operators ${\cal O}_{u,d}$ 
in the SUSY breaking sector
\begin{equation}
  {\cal L} = \int\!d^2\theta\,
    (\lambda_u H_u {\cal O}_u + \lambda_d H_d {\cal O}_d) + {\rm h.c.}
\label{eq:H-SUSY}
\end{equation}
Here, $\lambda_{u,d}$ are the renormalized couplings at the scale $M$. 
By rescaling the operators ${\cal O}_{u,d}$ we can always make the 
couplings $\lambda_{u,d}$ dimensionless, and we adopt this convention 
below.  Note that our discussion here applies to a very large class 
of theories -- the SUSY breaking sector can be strongly or weakly 
coupled, can contain single or multiple scales, and can lead to 
direct or indirect mediation of SUSY breaking.

After including the interactions in Eq.~(\ref{eq:H-SUSY}), the mass 
parameters in the Higgs sector receive a direct contribution from 
the SUSY breaking sector.  Assuming that the SUSY breaking sector 
does not have a special structure (such as an approximate PQ symmetry), 
the contribution is given by
\begin{equation}
  \mu \approx \lambda_u \lambda_d \frac{N_H}{16\pi^2} \Lambda_H,
\qquad
  B_\mu \approx \lambda_u \lambda_d \frac{N_H}{16\pi^2} \Lambda_H^2,
\label{eq:mu_b}
\end{equation}
\begin{equation}
  m_{H_{u,d}}^2 \approx \lambda_{u,d}^2 \frac{N_H}{16\pi^2} \Lambda_H^2,
\qquad
  A_{H_{u,d}} \approx \lambda_{u,d}^2 \frac{N_H}{16\pi^2} \Lambda_H,
\label{eq:mHu2_mHd2}
\end{equation}
where $N_H$ is the effective number of messenger fields coupled to $H_{u,d}$ 
in Eq.~(\ref{eq:H-SUSY}).  When the SUSY breaking sector is perturbative, 
$\Lambda_H$ is $F/M$ of these fields, and the sign of $m_{H_d}^2$ is, as 
required, positive in the simplest models (in which ${\cal O}_{u,d}$ are 
bilinears of messenger fields having generic $M$'s and $F$'s).  Generically, 
we expect $N \approx N_H$ and $\Lambda \approx \Lambda_H$, although they 
can easily differ by model dependent $O(1)$ coefficients.  Note that 
$m_{H_u,H_d}^2$ also receive the contribution of Eq.~(\ref{eq:GMSB}), 
so that their values are given by the sum of Eqs.~(\ref{eq:GMSB}) and 
(\ref{eq:mHu2_mHd2}).  Since the superparticle masses should not be 
much larger than a TeV to address the gauge hierarchy problem, the 
scales $\Lambda$ and $\Lambda_H$ are determined as
\begin{equation}
  \Lambda \approx \Lambda_H \approx O(10~\mbox{--}~100)~{\rm TeV}.
\label{eq:Lambda-range}
\end{equation}

For $\lambda_{u,d} \approx O(1)$, Eqs.~(\ref{eq:mu_b},~\ref{eq:mHu2_mHd2}) 
give $B_{\mu} \sim m_{H_{u,d}}^2 \gg \mu^2$.  This pattern has two 
problems.  One is that Eq.~(\ref{eq:minim-2}) does not allow a solution 
with sufficiently large $\tan\beta$, and the other is that the hierarchy 
between $m_{H_u}^2$ and $\mu^2$ leads to fine-tuning of $O(N_H/16\pi^2)$ 
or larger.  The first problem can be addressed by making $\lambda_u$ 
smaller, leading to $m_{H_u}^2 \ll B_\mu \ll m_{H_d}^2$, while the 
second problem by making $\lambda_d$ and/or $N_H$ larger, reducing 
the hierarchy between $m_{H_u}^2$ and $\mu^2$.  Note that the last two 
terms appearing in the RHS of Eq.~(\ref{eq:minim-1}), $m_{H_u}^2$ and 
$m_{H_d}^2/\tan^2\!\beta$, are always of the same order regardless of 
the values of $\lambda_{u,d}$.  Our proposal thus is to take
\begin{equation}
  \lambda_u \ll \lambda_d.
\label{eq:Bmu-sol}
\end{equation}
The simplest pattern of Eq.~(\ref{eq:muBmu-naive}) corresponds to 
taking $\lambda_d \approx 4\pi/\sqrt{N_H}$, the largest possible 
value, with $\epsilon \equiv \lambda_u \sqrt{N_H}/4\pi$.  The value 
of $\lambda_d$, however, need not be this large.  As stressed before, 
without eliminating the ``irreducible'' top-stop fine-tuning, there 
is little point in increasing $\lambda_d$ in order to reach the 
``fully natural'' pattern of  Eq.~(\ref{eq:muBmu-naive}).

There are several constraints on how large we can make $\lambda_d$ at 
$M$.  First of all, to have a very large value of $\lambda_d$, the SUSY 
breaking sector must be strongly coupled; otherwise $\lambda_d$ hits 
the Landau pole just above $M$.  Large $\lambda_d$ also strongly 
violates any parity invariance associated with hypercharge conjugation, 
leading to a possibility of introducing a too large Fayet-Iliopoulos 
(FI) contribution to $m_I^2$ at $M$.  Finally, large $\lambda_d$ 
implies large $m_{H_d}^2$, so that depending on the value of $\tan\beta$, 
it may lead to a tachyonic sbottom or stau.  These issues, however, 
are model dependent, and they can be addressed in explicit models 
with the outcome that $\lambda_d$ can in general take any value up 
to $\approx 4\pi/\sqrt{N_H}$~\cite{CFNV}.

Several mechanisms can produce the hierarchy of Eq.~(\ref{eq:Bmu-sol}). 
Since the hierarchy is not required to be larger than, say ${\cal O}(10)$, 
perhaps the simplest possibility is to assume that it arises as an 
accidental hierarchy of dimensionless numbers.  Alternatively, a dynamical 
realization can be found by assuming that the SUSY breaking sector is 
strongly coupled and approximately conformal over a range of scales 
above $M$.  In such a case, $\lambda_{u,d}$ are power-law sensitive to 
the anomalous dimensions of the operators ${\cal O}_{u,d}$, and a large 
hierarchy can naturally arise due to a small difference of these anomalous 
dimensions.  This scenario has a dual realization in 5D AdS space.  In 
that picture, the MSSM matter fields are localized on the UV brane, while 
the gauge and Higgs fields live in the 5D bulk and can directly couple to 
the SUSY breaking sector localized on the IR brane~\cite{Nomura:2004zs}. 
The hierarchy between $\lambda_u$ and $\lambda_d$ can then be explained 
by different localization of $H_u$ and $H_d$ in the 5th dimension.

{\bf Flavor and {\boldmath $CP$}.}
Before moving to phenomenology, let us shortly comment on the SUSY 
flavor and $CP$ problems. Since in our framework the SUSY breaking 
parameters are generated by gauge mediation and the direct Higgs 
couplings, they are automatically flavor universal at the scale $M$. 
The SUSY flavor problem is solved, provided that low energy radiative 
corrections do not induce large flavor violation.  One might worry 
that such corrections are present in our case, since $m_{H_d}^2$ and 
$\tan\beta$ are enhanced by powers of $\lambda_d/\lambda_u \gg 1$.  Through 
the RG equations, the large $m_{H_d}^2$ feeds in to the squark masses 
leading to $(m_Q^2)_{ij} \simeq \frac{1}{2} (m_D^2)^\dagger_{ij} \simeq 
-\frac{(y_d^\dagger y_d)_{ij}}{8\pi^2} m_{H_d}^2 \ln\frac{M}{m_{H_d}}$. 
However, in the SUSY field basis where $(y_{u,d})_{ij}$ are diagonal, 
the enhanced contribution to non-diagonal squark masses appears only 
in the mass-squared matrix of the left-handed up-type squarks.  It 
then follows that only nontrivial mass insertion parameters generated 
are $(\delta^u_{LL})_{ij}$.  Since the experimental constraints on 
these mass insertion parameters are rather weak, $(\delta^u_{LL})_{12} 
\simlt (10^{-2}\mbox{--}10^{-1})$, they do not lead to appreciable 
constraints on the theory.

We now consider the SUSY $CP$ problem.  Suppose that the operator 
${\cal O}_u$ (${\cal O}_d$) in Eq.~(\ref{eq:H-SUSY}) consists of a single 
term, so that there is only a single coupling $\lambda_u$ ($\lambda_d$). 
In this case, the phases of $\lambda_{u,d}$ can be completely absorbed 
into the phases of $H_u, H_d, Q$ and $L$.  Therefore, if the SUSY breaking 
sector preserves $CP$, all the parameters generated, $M_a$, $m_I^2$, 
$\mu$, $B_\mu$, $A_{H_u,H_d}$ are real in the basis where $\lambda_{u,d}$ 
are real, solving the SUSY $CP$ problem.  Since it is fairly simple to 
construct SUSY breaking models that preserve $CP$, we find this solution 
very attractive.  Note that in general $CP$ invariance in the SUSY 
breaking sector is not enough to solve the SUSY $CP$ problem; for example, 
the mechanism of~\cite{Giudice:1988yz} does not solve the problem even 
if the SUSY breaking sector preserves $CP$.  Here the structure of the 
couplings in Eq.~(\ref{eq:H-SUSY}) automatically leads to a solution 
to the SUSY $CP$ problem as long as the SUSY breaking sector does not 
violate $CP$.

{\bf Basic phenomenology.}
The most distinct phenomenological consequences of our framework follow 
from the fact that $m_{H_d}^2$ dominates all other soft terms.  First 
of all, the mass of the $CP$-odd neutral Higgs, $m_{A^0}^2 = 2|\mu|^2 
+ m_{H_u}^2 + m_{H_d}^2$, is much larger than the weak scale and is 
likely beyond the reach of the LHC.  Thus, the Higgs sector phenomenology 
corresponds to a special case of the so-called ``decoupling limit'' of 
the MSSM~\cite{Gunion:2002zf}.  For the scalar sector, we effectively 
see a one-Higgs doublet model at the weak scale.  On the other hand, 
the Higgsinos are relatively light since their masses are set by $\mu$. 
Small $\mu$ implies a fairly light chargino, and in the case $\mu < M_1$, 
the lightest neutralino is Higgsino-like.

The hierarchy $m_{H_u}^2 \ll m_{H_d}^2$ has also an important impact, 
via RG evolution, on the sfermion mass spectrum.  The large value of 
$m_{H_d}^2$ will have the unusual effect of turning on a sizable FI 
$D$-term for $U(1)_Y$, which leads to new contributions in the low 
energy superparticle spectra
\begin{equation}
  \delta m_I^2 \simeq Y_I \frac{3 g_1^2}{40\pi^2} 
    m_{H_d}^2 \ln\frac{M}{m_{H_d}},
\label{eq:FI-corr}
\end{equation}
where $Y_I$ represents the hypercharge.  In particular, the usual GMSB 
sum rules ${\rm Tr} Y m^2 = {\rm Tr} (B-L) m^2 = 0$ (with the trace 
running over one generation) are not satisfied; the traces are not 
even RG invariant.  Still, the relation ${\rm Tr} Y m^2 - \frac{5}{4} 
{\rm Tr} (B-L) m^2 = 0$ predicted in general GMSB~\cite{Meade:2008wd} 
is approximately RG invariant for the first two generations and does 
not receive a correction from the FI term.  It then follows that the 
sum rule $6m_Q^2 + 3m_U^2 - 9m_D^2 - 6m_L^2 + m_E^2 = 0$ is obeyed, 
while the traditional sum rule is modified to ${\rm Tr} Y m^2 \simeq 
\frac{g_1^2}{4\pi^2} m_{H_d}^2 \ln\frac{M}{m_{H_d}}$.  Once the detailed 
superparticle spectrum is known, the violation of this sum rule could 
point to large $m_{H_d}^2$, which may provide another hint that our 
scenario is realized.

For the third generation, the largeness of $m_{H_d}^2$ and $\tan\beta$ 
may significantly affect the RG evolution.  In particular, the left-handed 
squark $\tilde{q}_3$ and the right-handed sbottom $\tilde{b}$ receive 
the correction
\begin{equation}
  m_{\tilde{q}_3}^2 \simeq \frac{1}{2} m_{\tilde{b}}^2 \simeq 
    \frac{y_b^2}{8\pi^2} (m_{H_d}^2 + |A_{H_d}|^2) \ln\frac{M}{m_{H_d}},
\label{eq:sbottom}
\end{equation}
and similarly for the left-handed slepton $\tilde{l}_3$ and the 
right-handed stau $\tilde{\tau}$ with $y_b \rightarrow y_\tau$. 
In fact, for given values for the other parameters, this provides 
an upper bound on $\tan\beta$ from the experimental bounds on the 
masses of these particles, because $y_{b,\tau} \simeq (m_{b,\tau}/v) 
\tan\beta$.

{\bf Examples.}
In order to illustrate the features of low energy 
spectra, we have generated several example spectra using 
{\tt SuSpect}~\cite{Djouadi:2002ze}, and double-checked their 
consistency with {\tt SOFTSUSY}~\cite{Allanach:2001kg}.  When 
studying GMSB spectra one needs to take into account the fact that 
GMSB models are often more constrained experimentally than generic 
MSSM models, due to the characteristic $2 \gamma +\met$ signature 
arising from the next-to-lightest SUSY particle (NLSP) decaying into 
the gravitino, $\tilde{\chi}^0_1 \to \tilde{G} + \gamma$.  Tevatron 
searches for this final state set an upper bound on the total cross 
section involving superparticle production, which can imply bounds as 
strong as $m_{\tilde{\chi}^0_1} \geq 126~{\rm GeV}$ for the neutralino 
and $m_{\tilde{\chi}^+_1} \geq 231~{\rm GeV}$ for the chargino for 
a particular GMSB point~\cite{D0}.  More generally, this can be turned 
into a bound on the $\mu\mbox{-}M_1$ plane, as has been recently shown 
in~\cite{MRS}, which applies as long as the NLSP decays within the 
detector (which would generically be the case for low scale SUSY 
breaking as considered in this letter).

\begin{table*}
\begin{center}
\begin{tabular}{|c||c|c|c|c|c|c||c|c|c|c|c|c|c|c|c|}
\hline
 & \multicolumn{2}{c}{$\Lambda$} \vline & $\Lambda_H$ & 
 $\lambda_u$ & $\lambda_d$ & $\tan\beta$ & $m_{H_u}^2$ & $\mu$ & 
 $m_{h^0}$ & $m_{A^0}$ & $m_{\tilde{\chi}_1^0}$ & $m_{\tilde{\chi}_1^+}$ & 
 $m_{\tilde{t}_1}$ & $m_{\tilde{t}_2}$ & $m_{\tilde{\tau}_1}$ \\
\hline \hline
 M1 & \multicolumn{2}{c}{60} \vline & 18 & 0.50 & 2.80 & 8.01 & 
 $(521)^2$ & 160 & 115 & 4030 & 150 & 159 & 1360 & 1520 & 353 \\
\hline
 M2 & \multicolumn{2}{c}{40} \vline & 20 & 0.28 & 2.08 & 10.13 & 
 $(302)^2$ & 196 & 113 & 3330 & 164 & 188 & 923 & 1040 & 192 \\
\hline
 M3 & 19 & 45 & 10 & 0.58 & 3.05 & 4.67 & 
 $(483)^2$ & 221 & 101 & 2520 & 167 & 209 & 431 & 607 & 256 \\
\hline
 M4 & 19 & 50 & 10 & 0.27 & 3.93 & 8.08 & 
 $(356)^2$ & 182 & 105 & 3160 & 155 & 176 & 432 & 632 & 293 \\
\hline
\end{tabular}
\end{center}
\caption{Four sample spectra illustrating the features of our approach 
 to the $\mu\mbox{-}B_\mu$ problem, obtained using {\tt SuSpect}.  The 
 first 5~columns define the input parameters.  At the scale $M$, which 
 we take to be twice the largest of $\Lambda$ and $\Lambda_H$, the gaugino 
 and scalar masses are given by Eq.~(\ref{eq:GMSB}), while $m_{H_{u,d}}^2$ 
 and $A_{H_{u,d}}$ also receive contributions from Eq.~(\ref{eq:mHu2_mHd2}) 
 with the coefficients $+1$ (we set $N = 4$, $N_H = 1$ in all cases). 
 For the M3 and M4 points we use the separate triplet and doublet scales 
 $\Lambda_3$ (left) and $\Lambda_2$ (right) as described in the text. 
 The $\mu$ and $B_\mu$ parameters are computed from $\tan\beta$ and the 
 electroweak scale $v$ using the EWSB condition, whose values at $M$ are 
 consistent with Eq.~(\ref{eq:mu_b}) up to an $O(1)$ factor.  The remaining 
 9~columns are sample output parameters; $m_{H_u}^2$ and $\mu$ represent 
 their values at the weak scale.  Note that the spectrum is invariant 
 under the rescaling of the input parameters $\lambda_{u,d} \rightarrow 
 \lambda_{u,d}/\alpha , N_H \rightarrow \alpha^2 N_H$.
\label{tab:examples}}
\end{table*}
Four explicit example spectra are presented in Table~\ref{tab:examples}. 
The first example point (M1) is the most conservative: it uses minimal 
GMSB with very heavy stops (over $1.4~{\rm TeV}$), where the Higgs boson 
is sufficiently heavy without any extra source for the Higgs quartic 
coupling.  In this case, there is still a significant (``irreducible'') 
fine-tuning arising from the top-stop contribution to $m_{H_u}^2$.  The 
second example (M2) has a somewhat lighter stop (of order $1~{\rm TeV}$), 
with the gaugino and sfermion masses still following the pattern of 
minimal GMSB.  The fine-tuning due to the top-stop contribution is reduced, 
but one may need to have a small extra contribution to the Higgs quartic 
coupling.  In the last two examples, we deviate from minimal GMSB, and 
``squash'' the sfermion spectrum by assuming different $F/M$ values 
for the triplet and doublet messengers, using the notation $\Lambda_3 
= (F/M)_3$, $\Lambda_2 = (F/M)_2$, and $\Lambda_1 = \sqrt{(2/5)\Lambda_3^2 
+ (3/5)\Lambda_2^2}$.  In this case the stop mass is lowered to around 
$400~{\rm GeV}$, and the fine-tuning associated with the ordinary little 
hierarchy problem is strongly softened (the fine-tuning in this point 
is of order $\approx 10\%$). However, since the top-stop contribution 
to the Higgs mass is now reduced, a sizable extra contribution to the 
Higgs quartic coupling must be present.  In all four cases, $m_{H_u}^2$ 
is positive at the weak scale (although one can find examples where 
it is negative), $\mu$ is small, and $m_{A^0}$ (which is roughly 
equal to $m_{H_d}$) is much larger than TeV.  The presented examples 
should be considered as a proof of principle that our approach to 
the $\mu\mbox{-}B_\mu$ problem can be consistently realized.

{\bf Summary.}
We have shown that it is {\em not} necessary to reduce $B_\mu$ in 
GMSB models.  Successful EWSB can be obtained with the hierarchy of 
Eq.~(\ref{eq:pattern}), which can naturally arise if the Higgs fields 
are directly coupled to the SUSY breaking sector.  EWSB can be achieved 
without turning $m_{H_u}^2$ negative.  This solution elegantly addresses 
the $\mu\mbox{-}B_\mu$ problem of GMSB, without reintroducing the SUSY 
flavor problem.  The SUSY $CP$ problem can also be addressed.  Specific 
phenomenological predictions include a decoupled second Higgs doublet, 
relatively light Higgsinos, large violation of the traditional mass 
sum rule, and large corrections to the sbottom and stau masses.  The 
scenario studied in this letter can be applied to large classes of 
GMSB models, and is consistent with gauge coupling unification if the 
couplings $\lambda_{u,d}$ are not too large or are asymptotically free. 
We will further elaborate on various issues associated with the present 
scenario in the upcoming paper~\cite{CFNV}.

{\bf Acknowledgments.}
We thank K.~Agashe, N.~Arkani-Hamed, R.~Dermisek, Z.~Komargodski, 
P.~Meade, M.~Papucci, M.~Reece, D.~Shih, P.~Slavich, S.~Su, and S.~Thomas 
for useful discussions and to J.~L.~Kneur for sending us a new version 
of {\tt SuSpect}.  C.C., A.F. and T.V. thank the KITP at Santa Barbara 
and C.C., Y.N. and T.V. thank the Aspen Center for Physics for their 
hospitality.  The work of C.C. is supported in part by the NSF under 
grant PHY-0355005 and by a US-Israeli BSF grant.  A.F. is partially 
supported by the European Community Contract MRTN-CT-2004-503369.  The 
work of Y.N. is supported in part by the NSF under grant PHY-0555661, 
by a DOE OJI award, and by the Alfred P. Sloan Research Foundation. 
The work of T.V. is supported in part by the DOE grant DE-FG02-90ER40542.

\end{document}